\documentclass[vecphys]{svmult}

\usepackage{makeidx}         % allows index generation
\usepackage{graphicx}        % standard LaTeX graphics tool
\usepackage{multicol}        % used for the two-column index
\usepackage[bottom]{footmisc}% places footnotes at page bottom
\makeindex             % used for the subject index
\begin{document}

\title*{Short Gamma Ray Bursts: marking the birth of black holes from
coalescing compact binaries}
\author{Davide Lazzati\inst{1} \and
Rosalba Perna\inst{1}}
\institute{JILA, University of Colorado, 440 UCB, Boulder, CO
80309-0440, USA
\texttt{lazzati,rosalba@colorado.edu}
}
\maketitle

\section{Introduction}
As soon as the catalog of gamma-ray bursts (GRBs) detected by the
BATSE (Burst And Transient Source Experiment) had enough events to
allow a statistical study, it was discovered that GRB light curves
could be separated in two families \cite{kouv93}. Long GRBs are
characterized by a duration of more than 2 seconds and a somewhat soft
spectrum. Short GRBs, on the other hand, are characterized by a
duration of less than 2 seconds and a harder spectrum.

Not much more could be said in the BATSE era, due to the lack of
precise localizations and impossibility of long wavelength follow-up
that plagued both the long and short GRB populations. With the launch
of the Italian-Dutch satellite BeppoSAX, the situation for long GRBs
changed dramatically. X-ray, optical and radio afterglows were
discovered \cite{cost97,vanp97,tayl97} to follow the prompt
phase. Spectroscopy revealed that the GRBs lie at cosmological
distances and that involve explosion energies similar to core collapse
supernovae \cite{metz97,kulk99}. Evidence of beaming and association to
massive stars emerged \cite{rhoa99,sari99} leading to the now widely
accepted scenario of long GRBs as collimated relativistic outflows
associated to Type Ib/c supernova explosions
\cite{stan03,hjor03}. Unfortunately BeppoSAX was non optimally
designed to detect short GRBs and none of the above information was
available for the short bursts that remained elusive and
mysterious. The only advance came from the discovery that short GRBs
also have longer wavelength emission on longer timescale
\cite{lazz01}. Such discovery was however made on a stacked light
curve from past events, and did not allow for any follow-up
observation. A general consensus was reached in those years that short
GRBs could be associated to the merger of compact binary systems
\cite{eich89}, based on a theoretical desire more than on any robust
evidence.

More recently, thanks to the HETE-2 satellite and to Swift, afterglow
observations have been performed also for the class of short GRBs,
measuring their redshift and energetics, and finally giving some
observational corroboration to the idea that they originate from
binary mergers \cite{vill05,gehr05}. Not everything has been clarified,
though, the main problem being now the one of defining what is a long
and what is a short bursts, since the two classes seem to have a gray
area between them with bursts sharing a complex set of properties. In
this review, we critically present the new discoveries made with HETE-2
and Swift, the theoretical advances that were made possible by those
discoveries and the still debated issues and future perspectives. This
paper is divided in two main sections, the first observational and the
second theoretical.

\section{Observations}

\subsection{Pre-Swift era}

\begin{figure}
\includegraphics[width=1.0\columnwidth]{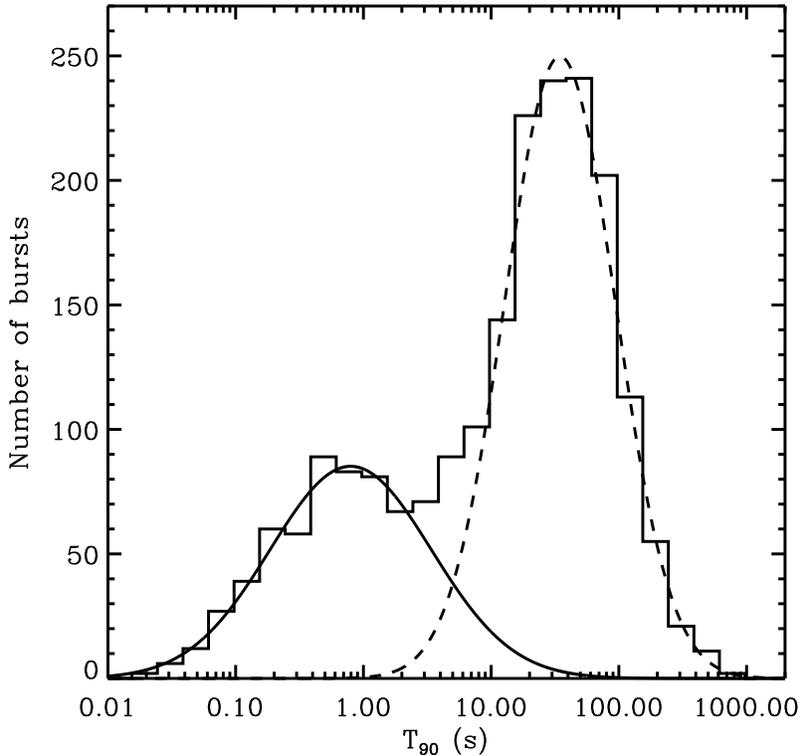}
\caption{Distribution of the durations of 2041 BATSE GRBs. Gaussian
fits for the short (solid line) and long (dashed line) populations are
overlaid.}
\label{fig:t90}
\end{figure}

Observations performed in the pre-Swift (and HETE-2) era are mainly
those performed with BATSE and Konus. Figure~\ref{fig:t90} shows a
histogram of the $T_{90}$ distribution of 2041 GRBs detected by
BATSE. The $T_{90}$ is the time interval during which the GRB emits 90
per cent of the total fluence. The solid and dashed lines show the
Gaussian fit to the distribution for the short and long bursts,
respectively. Even though the bimodality is clear, it is also clear
that the two populations are not entirely separated, since a
considerable tail of short burst population (about 25 per cent)
extends beyond $T_{90}>2$~s, the traditional dividing line.

\begin{figure}
\includegraphics[width=1.0\columnwidth]{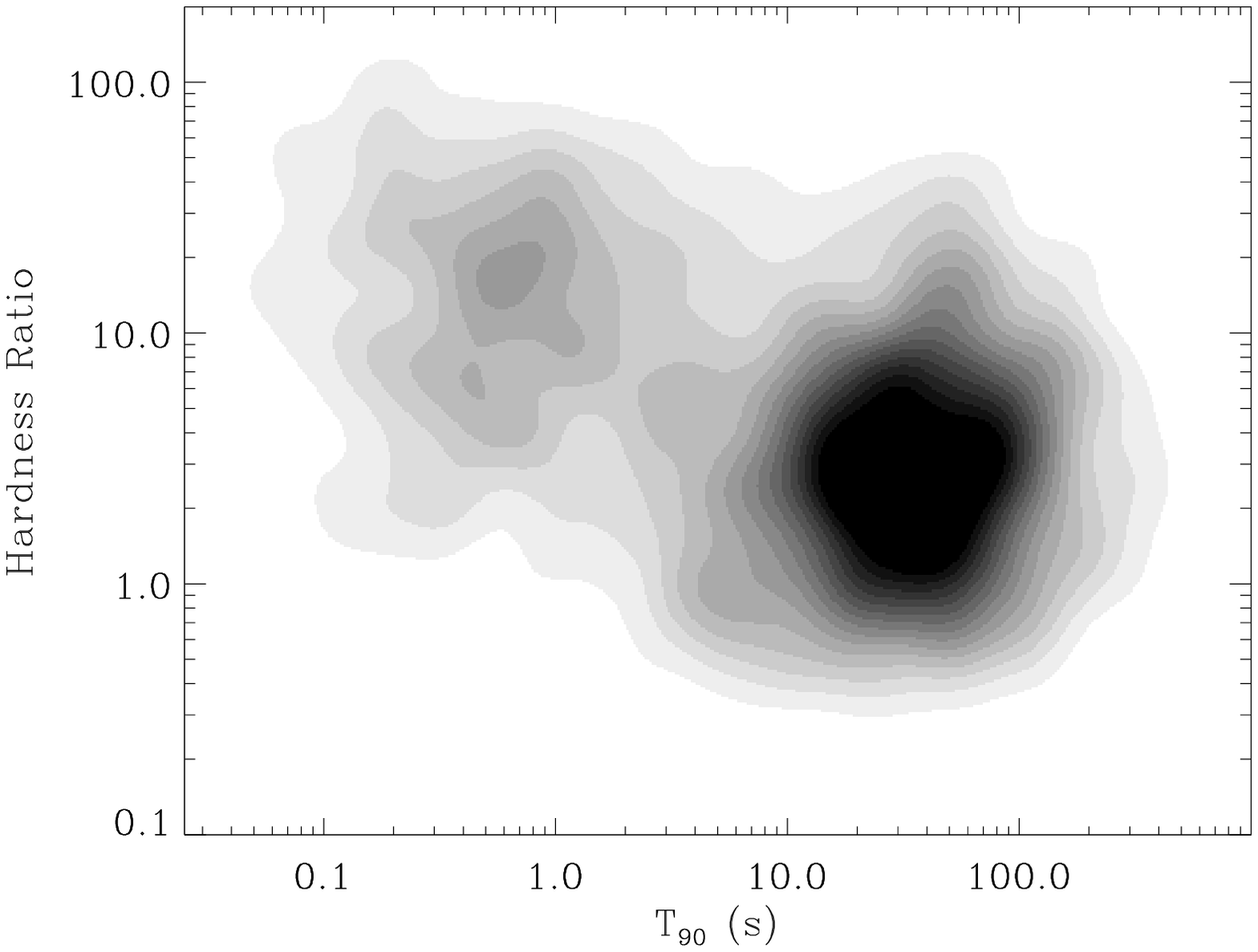}
\caption{Hardness-duration distribution of BATSE GRBs.}
\label{fig:spex}
\end{figure}

Additional separation between the two classes is provided by the
spectral analysis of the light curves. Figure~\ref{fig:spex} shows in
greytones the two dimensional distribution of BATSE GRBs in the
hardness-duration plane. The hardness is defined as the ratio of
counts in the high energy channels (3 and 4) over the counts in the
low energy channels (1 and 2). Even though short bursts appear to have
a systematically harder spectrum, the two population have a sizable
overlap.

The origin of the spectral difference has been analyzed in detail by
Ghirlanda et al. \cite{ghir04}. They performed spectroscopy of a sample
of 36 bright short bursts comparing the results of several spectral
models. They conclude that the spectra of short bursts are successfully
fit by a single power-law with an exponential high energy
cut-off. They also find that short GRBs have harder spectra due to
steeper low energy slopes of the power-law rather than due to a larger
perk frequency. It is worth reminding that long GRBs are usually fit
wit a smoothly broken power-law model or Band function \cite{band93}.

Nakar \& Piran \cite{naka02} analyzed the light curves of a sample of
short GRBs from BATSE. They find that, when high resolution data are
available, short GRB light curves can be resolved into the
superposition of many pulses, with statistical properties analogous to
those of the long GRBs. This may indicate that the same dynamical and
dissipation processes are powering the light curves, even though such
processes are still far from being understood.

\begin{figure}
\includegraphics[width=1.0\columnwidth]{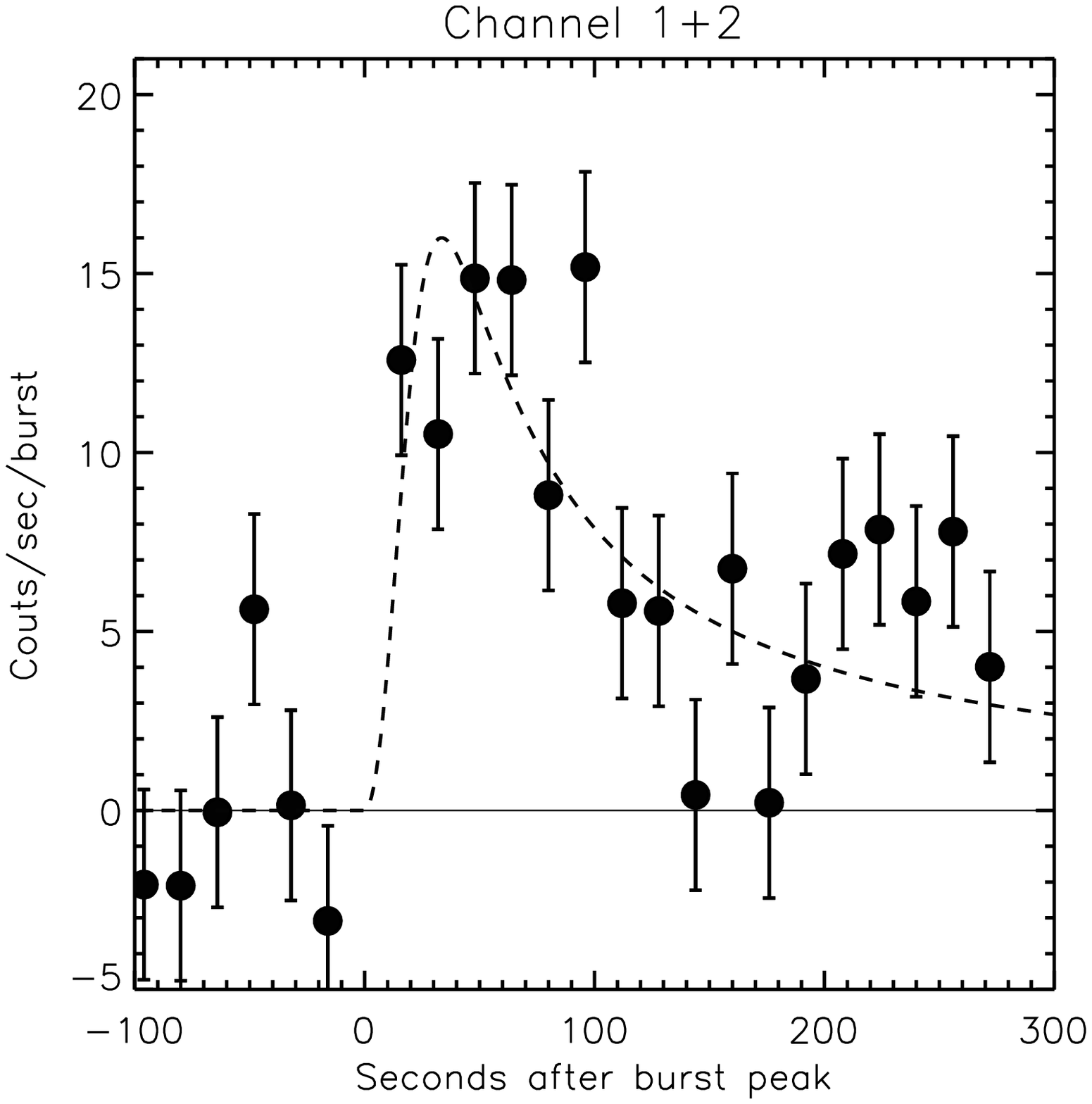}
\caption{Light curve of the excess emission found by Lazzati et
al,~\ref{lazz01} in the composite light curve of 76 bright BATSE short
GRBs. The dashed line is a best fit afterglow model.}
\label{fig:l01}
\end{figure}

The quest for short GRB afterglows went on for all the BATSE and
BeppoSAX era with very limited results. No afterglow of an individual
short GRB was ever found. Lazzati et al. \cite{lazz01} stacked the
background subtracted light curve of the 76 brightest short BATSE GRBs
looking for an evidence of afterglow in the hard X-rays. They found
(see Fig.~\ref{fig:l01}) that there is indeed an excess soft component
following the prompt emission of short GRBs lasting approximately 100
seconds. They showed that this component is consistent with being of
afterglow origin. As we will see below, it later emerged from Swift
observations that this component is more likely residual activity from
the central engine.

\subsection{The Swift era}

Many of the riddles of short GRB astrophysics have been solved in the
Swift era, with a noticeable contribution from the HETE-2
satellite. Due to the characteristics of the BAT detector, Swift
observations did not increase our understanding of the prompt emission.

In the spring and summer of 2005, after years of struggle, afterglow
of short GRBs were finally detected in X-rays, optical and radio
wavelengths \cite{gehr05,vill05,cast05,fox05,hjor05,covi06,berg05},
bringing the short bursts in the afterglow era.

\subsubsection{Afterglows}

\begin{figure}
\includegraphics[width=1.0\columnwidth]{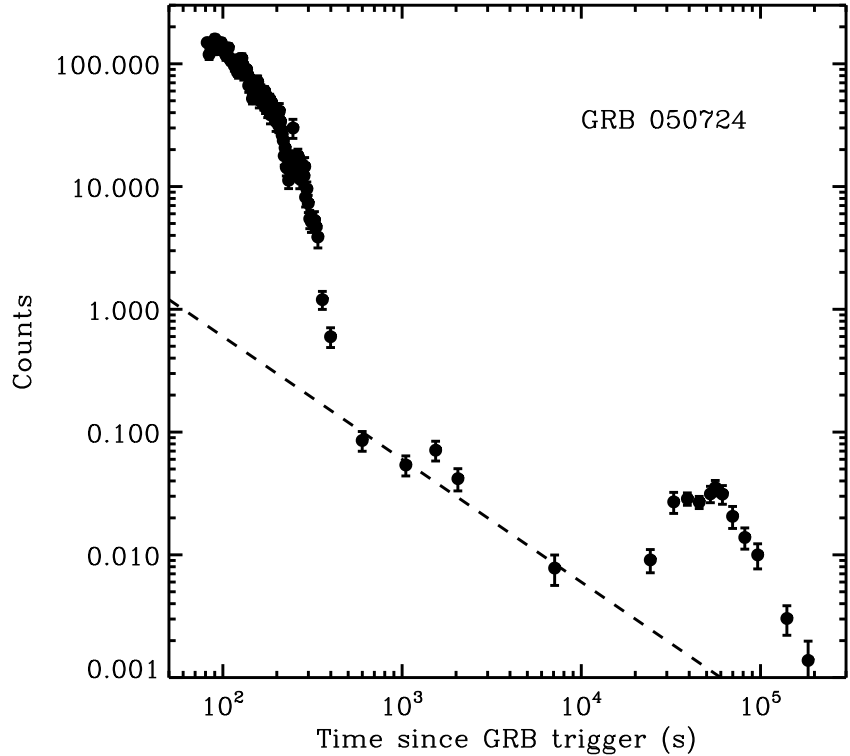}
\caption{Swift X-ray afterglow of GRB~050724 \cite{camp06}. The dashed
line shows a $t^{-1}$ power-law for comparison purposes.}
\label{fig:lcx}
\end{figure}

At first sight, afterglows of short GRBs are similar to the afterglows
of long GRBs. They are fainter, but qualitatively
analogous. Figure~\ref{fig:lcx} shows the X-ray afterglow of
GRB~050724 as observed by the Swift XRT \cite{camp06}. The afterglow
shows an initially very bright phase, followed by a sharp decline, a
possible $t^{-1}$ power-law decay and a flare at late times. The
afterglow is overall very faint. This is supposed to be due to a
combination of an isotropic equivalent energy smaller than that of
long bursts and of a low density interstellar medium, down to
$n\sim10^{-5}$~cm$^{-3}$ \cite{naka07}.  Such low densities are
thought to be an indication of the binary merger origin of short
bursts, since are of the order of magnitude of what expected in the
intergalactic medium.

The initial bright phase is likely the one detected on BATSE data
\cite{lazz01} and initially interpreted as the regular afterglow. It
is now believed to be a sign of continued activity of the central
engine (see below). X-ray flares are also common. Differently from the
X-ray flares detected on top of long GRB afterglows, the flares so far
detected on long short GRBs have long time scale ($\delta{t}/t\sim1$)
\cite{lazz07}. Their origin is not clear, a possible detection of
rapid variation was reported for GRB~050709 \cite{fox05}, but is
inconclusive. Since rapid variations are associate to late time
activity of the engine \cite{lazz07}, their confirmation would be of
great relevance for our understanding of the short GRB engine physics.

It is still debated whether short GRB afterglows show conclusively the
presence of beaming of the short GRB fireballs. The question is
relevant since the total energy of the explosion depends on the
beaming factor. In few cases, a jet break has been claimed. A good
example is that of GRB~051221 \cite{sode06} where a simultaneous break
in the optical and X-ray light curves was detected approximately 5
days after the explosion. This is however an isolated case and in most
case only an lower limit to the beaming can be obtained
\cite{grup07}. It seems fair to conclude that, even if beamed, short
GRBs are less beamed than long GRBs, for which opening angles of few
degrees are routinely measured \cite{ghir04a}. This observations
supports the idea that short GRBs are associated to the merging of
compact objects, for which hydrodynamical collimation has a much
lesser role than in massive stars \cite{lazz05}.

Intensive searches for a supernova component have been performed in
several short GRBs. Due to their moderate redshift, the searches are
very sensitive, being able to detect the presence of a supernova bump
even if the supernova was 10 times fainter than the faintest observed
supernova \cite{fox05,bloo06,covi06,sode06}. The lack of any detection
strongly favors a different progenitor for short and long GRBs.

\subsubsection{Host galaxies}

A debated and important aspect of short GRB research has been that of
the identification of the host galaxy. It is important for at least
two reasons. First, the identification of a host galaxy is usually the
only way to obtain a redshift for short GRBs. Second, host galaxies
can provide important clues to the nature of the progenitor of the
short bursts and on the physics of their afterglow emission.

Unlike long GRBs, short bursts provided a harder challenge to
astronomers for the identification of the host galaxy. While long
bursts explode on top of the brightest region of galaxies, making the
identification unquestionable \cite{fruc06}, short GRBs explode in
anonymous regions in the outskirt of galaxies, sometimes having more
than one galaxy within their error circles.

The first well-localized short GRB offers a good example of the
situation \cite{hjor05a,pede05,gehr05}. The X-ray error box was large
enough to contain a low-redshift elliptical galaxy, member of a
cluster, and several high redshift faint objects, analogous to the
host galaxies of long GRBs. The implication of the choice of host
were relevant. If the nearby elliptical was the host, short GRBS would
be associated to non star forming objects, they would explode in the
outskirts of the host - if not outside of them - and would be low
redshift events, involving several orders of magnitude less energy
than long events. On the other hand, if the progenitor was one of the
high redshift objects, short GRBs could be analogous to the long ones.
Circumstantial evidence favoring a low redshift origin of short bursts
was emerging in the meantime, showing that the location of BATSE short
GRBs correlates with local bright galaxies \cite{tanv05} or with
clusters \cite{ghir06}.

With more short GRB localizations and improvement in the error boxes
there is now very little doubt that the population of short GRB host
galaxies is markedly different from those of long bursts. Short GRB
hosts seem to be a far less homogeneous sample than those of long
ones. With the caveat that there may be some misidentifications, given
that short GRBs do not explode in the brightest parts of their hosts
but rather in their outskirts. Long GRB hosts are usually dwarf
starbursting galaxies at moderate to high redshift, always found in
the field. Morphologically, short GRB hosts can be of any kind, from
early type ellipticals to late type spirals, and are found both in the
field and inside clusters. In most cases, short GRB hosts have low
star formation, $<1M_\odot/$y$^{-1} (L_\star/L)$, or about one
hundredth of the star formation rate in long GRB hosts
\cite{chri04}. The two populations are different at a very high
confidence level \cite{goro06} and at least in some cases, evidence of
an old population of star was found in the host
\cite{covi06,sode06}. Approximately 20 per cent of Swift short bursts
are associated to clusters of galaxies \cite{berg06}, in agreement
with the fraction of stellar mass contained in such systems. Short
GRBs seem therefore to be better unbiased samples of the stellar
population in the low-moderate redshift universe than long ones.

\subsubsection{Redshift distribution}

\begin{figure}
\includegraphics[width=1.0\columnwidth]{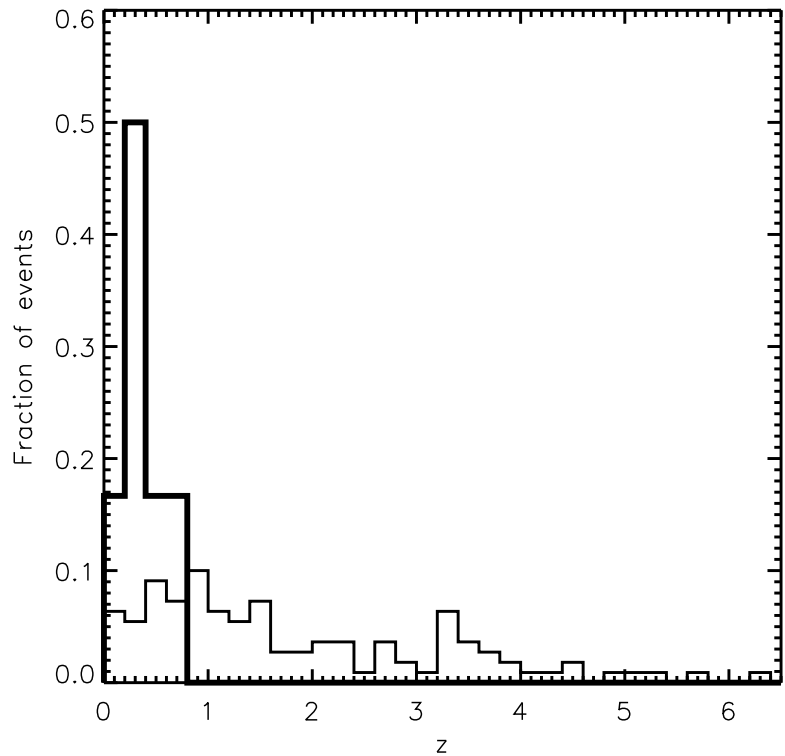}
\caption{Redshift distribution of short (thick line) and long (thin
line) GRBs.}
\label{fig:z}
\end{figure}

The redshift of short GRBs is always measured as the redshift of the
putative host galaxy and is therefore less robust than the long GRB
redshifts, that are often measured both in absorption on the afterglow
spectrum and in emission in the host galaxy. Figure~\ref{fig:z} shows
the short GRB redshift distribution with a thick line histogram. Even
though some of the values can be debated, it is clear that the
redshift distribution of short GRBs is centered at much smaller
redshifts than that of long GRBs. The long GRB redshift distribution
is shown with a thin histogram for comparison and extends beyond
redshift 6 \cite{tagl05,hais06,kawa06}.

\subsubsection{Short or not short?}

We have discusses so far a consistent body of observational evidence
coherently supporting the idea that short and long GRBs have different
physical origin and are two well separated populations. Long GRBs are
long in $\gamma$-rays and soft, have bright afterglows with jet breaks
and supernova bumps, they explode in star forming regions inside high
redshift dwarf irregular star forming galaxy. Short burst, on the
other hand, are short and hard in $\gamma$-rays, have dim afterglows
and lack any evidence of an associated supernova, explode in
low-intermediate redshift galaxies of all kind, with little sign of
star formation.

Such an idyllic interpretation framework was suddenly shaken last
summer, when two puzzling bursts were detected: GRB~060505 and
GRB~060614 \cite{dell06,fynb06,gal-06}. They appeared to be long in
the $\gamma$-ray properties, but had faint afterglows and no sign of
an associated supernova, down to very stringent limits. Possible
interpretations range from short bursts with a particularly bright
X-ray early afterglow \cite{zhan07} to long bursts with a non-massive
star progenitor \cite{gal-06}, or associated to a SN explosion that
did not produce large quantities of $^{56}$Ni \cite{tomi07}. The
puzzle becomes even more intricate if secondary indicators are
used. GRB~060614 is found to be consistent with the Amati correlation
\cite{amat07}, like a long duration GRB, but to share the short GRB
properties in terms of spectral lags (no significant lag was detected)
\cite{norr00}.

Even though it is definitely premature to give up a classification
successful for decades based on only two events, such observations
should clearly ring a warning bell. It is obvious that if short GRBs
are followed by relatively bright early X-ray afterglows
\cite{lazz01,camp06}, the distinction between short and long events
based only on their duration is band dependent and doomed to
fail. However, a new robust scheme has not appeared, so far. Spectral
lag are a promising tool, but they are still theoretically unexplained
and may well not be associated to the physics of the progenitor,
something that seem to be clearly different between the two classes. A
classification based on the host galaxy properties is also a
possibility that could be considered, with the problem that not all
GRBs have host galaxy detections and so many events would become
unclassified. It is clear that more events of difficult classification
have to be detected before we can establish a new, successful,
classification scheme.

\subsection{SGRs}

A considerable excitement was caused by the detection of a giant flare
from the soft gamma repeater (SGR) SGR~1806-20
\cite{palm05,tera05,came05,mere05}. The initial phase of the flare is
a bright spike lasting less than one second, with a blackbody spectrum
and enough energy to be detected by BATSE out to a distance of
approximately 50 Mpc. It is followed by a pulsed X-ray tail that would
be undetectable with any present instrumentation for an extragalactic
flare. These properties suggested that a significant fraction of BATSE
short GRBs could indeed be giant flares from SGRs in the local
universe, out to the Virgo cluster of galaxies. A large effort was
attempted to constrain this fraction and possibly identify the
extragalactic SGRs in the BATSE sample. The search was based on
positional coincidence with nearby galaxies
\cite{naka06,ofek06,popo06}, on the spectral properties of the prompt
emission \cite{lazz05a}, and on the presence of an oscillating
tail. None of these searches found any suitable candidate, implying a
fraction of at most 15 per cent of BATSE short GRBs being SGR flares
in the local universe. This limit is only marginally consistent with
the Galactic SGR rate, and requires the SGR1806-20 flare to be an
exceptional event. Of course, caution should always be considered when
statistics is based on a handful of events.

\section{Theory}

\section{Binary mergers as progenitors of short GRBs}

The leading candidates as progenitors of short GRBs are mergers of
NS-NS binaries \cite{pacz86,good86,eich89,nara92} and BH-NS binaries
\cite{nara92,moch95}. Two compact objects in a binary are bound to
eventually merge due to the emission of gravitational wave radiation
that causes a loss of energy resulting in a gradual shrinking of the
orbit.  For typical binary parameters, binaries are expected to
coalesce within a Hubble time, as discussed in more detail in the
following.  From an energetic point of view, there is enough
gravitational binding energy that is liberated during the merger to
power the GRB and its subsequent afterglow. The engine that helps
channeling this energy into a relativistic flow is believed to be,
like in the case of long GRBs, an hyper-accreting accretion disk.
However, while in the case of long bursts (which are thought to be
associated with the collapse of massive stars), the event duration is
set by the collapse time of the star envelope (on the order of several
tens of seconds) that keeps on feeding the disk, in the case of a
binary merger the activity phase is set by the viscous timescale of
the disk, which is a fraction of a second, the right order of
magnitude needed for the power source of short bursts.

From an observational standpoint, double neutron stars are known to
exist from direct observations in the Galaxy.  Well-known examples are
the Hulse-Taylor binary and the binary pulsar \cite{burg03}. On the
other hand, compact-object binaries where one of the components is a
black hole have not been observed so far. However, theoretical
modeling of binary evolution predicts that binaries with black holes
are expected (e.g. \cite{brow95, port98, frye98, belc02}). It should
be noted that, even if BH-BH binaries are also predicted as a possible
end state of binary evolution, however they are not expected to give
rise to a GRB when they merge. This is because, in this scenario,
there would be no fuel to power the accretion disk that could then
provide the source of energy for the GRB (see \S??).

In order for mergers of compact objects to be progenitors of short
bursts, a fundamental condition is that the merger event rate be comparable
to that of the bursts. With the growing number of short bursts with
detected redshift, a comparison can be made not only for the
overall (i.e. redshift-integrated) event rate, but also for its 
redshift evolution. As the sample grows, the calibration of
the cosmic rate evolution, in combination with observations of the
burst rate as a function of the galaxy type, can allow one to further
constrain various types of binary models. 

The main source of guidance for the expected characteristics of
a population of bursts associated with the coalescence of two
compact objects is provided by population synthesis calculations.
In the following, we will describe the main assumptions, 
calculation methods, and results from these calculations (\S??), 
with a special emphasis on their specific application to short
bursts (S??). Finally (\S??), we will discuss the highlights of 
numerical calculations that simulate the final moments of the binary,
and the formation of the accretion disk that eventually leads to the GRB.

\subsection{Binary evolution - theoretical modeling}

In the last few years, several groups \cite{port98,belc02} have
developed population synthesis calculations. These simulations,
starting from some assumptions regarding the binary star population,
track the evolution of stars both individually and in relation to the
companion in the binary system; in output, they are able to predict
the fraction of systems that end up with a certain type of compact
objects, the merger time of each binary, and hence the cosmic event
rate.  In the following, we describe in more detail the highlights of
the population synthesis calculations.

The evolution of each star in the binary system is computed given its
zero-age main-sequence mass and its metallicity. The code tracks all
the stages from the main sequence to the red giant branch to the core
helium burning and the asymptotic giant branch. At each stage of
evolution, the basic stellar parameters (radius, luminosity, stellar
mass, core mass) are determined. While each star is evolved depending
on its initial parameters, a number of effects that influence the
binary orbit (e.g. mass and angular momentum losses due to stellar
winds) are taken into account. At every evolutionary time step, the
codes check for possible binary interactions. If any of the components
fills its Roche lobe, then the resulting mass transfer is computed,
and so the eventual resulting mass and angular momentum losses. If the
binary survives the mass transfer event (i.e. both stellar components
fit within their Roche lobe), then the evolution of the binary keeps
on being followed.  The calculation for each star ends at the
formation of a stellar remnant: a white dwarf, a neutron star or a
black hole. If the remnant is born with a supernova explosion, the
effects of the supernova kicks and mass loss on the binary orbit are
computed.  Finally, once a binary consists of two remnants, its merger
lifetime is calculated, i.e. the time until which the components merge
due to emission of gravitational radiation and the consequent orbital
decay.

Early population synthesis studies \cite{port98,frye99} found that,
for the NS-NS binary systems, merger times are generally long, $t_{\rm
merg}>~0.1-1$ Gyr.  Smaller timescales were found \cite{tutu94} as a
result of the assumption that the secondary star, once it becomes a
low-mass helium-rich star, can initiate an extra mass transfer
phase. More recently, simulations \cite{belc02} identified a new
population of coalescing NS-NS binaries, which merge on a timescale
which is much shorter than that of the 'classical'' population.  These
very short timescales, $t_{\rm merg}~0.001-1$ Myr, are the result of
allowing both the primary and the secondary star to initiate an extra
mass transfer phase. Furthermore, the inclusion of natal kicks in the
simulations \cite{belc02} further contributes to decrease the merger
timescales. Accounting for natal kicks, in fact, results in the
disruption of the widest binaries and in an eccentricity gain for the
remaining ones, which further reduces their lifetimes. This
subpopulation of relatively short-lived NS-NS binaries was found to be
the dominant channel, making up about 80\% of the total population.

On the other hand, all population synthesis calculations agree on the
typical distribution of merger times of the NS-BH systems, which is
found to be similar to that of the ``classical'' population of NS-NS
binaries, i.e. $t_{\rm merg}>~0.1-1$ Gyr.

\subsection{Theoretical predictions for the observational properties of
short bursts due to binary mergers}

{\em a) Rates} The distribution of merger times obtained from
population synthesis calculations, combined with a prescription for
the cosmic star formation rate, allows one to estimate the rate of
binary mergers throughout the lifetime of the Universe.  Since the
simulations by Belczynski et al. \cite{belc06} include the population
of tight NS-NS binaries (i.e. the short-lived population), the
predicted cosmic redshift rates are different for the population of
NS-NS and of NS-BH binaries. An example is shown in Figure 1, with
data from the simulations by Belczynski et al. \cite{belc06}. Those
calculations used as star formation rate which peaks at a redshift of
about 3 and has only a mild decline up to redshifts of about 5.  Due
to the merger time delays, however, the predicted binary merger rate
peaks at a remarkably lower redshift.  The larger the time delay, the
larger the shift to lower redshifts of the binary merger event rate
compared to that of the underlying star forming rate. This is
especially emphasized in Figure 1, where the rates due to the fraction
of (both NS-NS and NS-BH) long-lived binaries ($t>100$ Myr) are shown
together with the overall rates. Whereas the assumed star-formation
rate remains constant up to a redshift of $\sim 5$, the rate of the
long-lived component drops substantially. For the overall binary
merger rate, if the assumed SFR peaks at a redshift of $\sim 3$, the
peak is found in the range $\sim 1-2$.

\begin{figure}
\includegraphics[width=1.0\columnwidth]{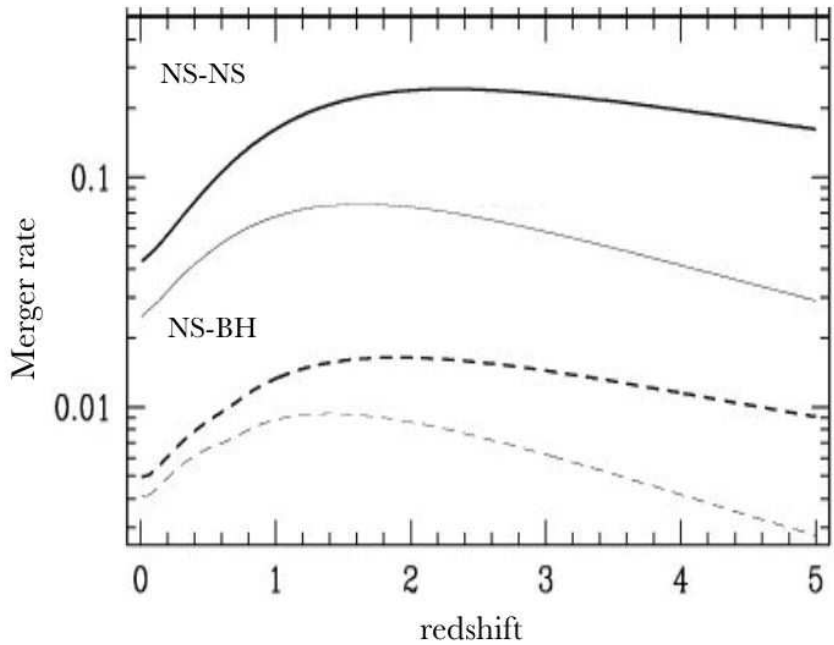}
\caption{{\rm Thick lines}: the inferred merger rate as a function of
redshift for the NS-NS and BH-NS mergers.  {\em Thin lines}: the
contribution from the long-lived ($t>100$ Myr) population.  Rates are
in arbitrary units. Data from the simulations of Belczynski et
al. \cite{belc06}.}
\label{rates}
\end{figure}

Another method of estimating the binary merger rates starts from a
count of the observed number of NS-NS binaries in the Milky Way. This
number is then corrected to account for the completeness of the survey
and transformed into a local rate by weighing in the estimated
lifetimes of the observed systems.  Up to date, the latest
calculations of the Galactic merger rates \cite{kalo94} have yielded a
value in the range $1.7\times 10^{-5}$ to $2.9\times 10^{-4}$
yr$^{-1}$ at the 95\% level.

{em b) Offsets from host galaxies, densities of the circumburst medium,
afterglow brightnesses and galaxy types}

Since the first studies on GRBs, it was clear that the location of a
burst within a galaxy, and hence its environment, would provide
important clues to the properties of the progenitor itself. In the
past decade, a number of groups \cite{frye99,bloo99,pern02,belc06}
have performed statistical studies of the distribution of the binary
merger sites in a galaxy.  Generally speaking, progenitors with very
short lifetimes are typically expected to produce GRBs close to their
place of birth. However, if the progenitor is endowed with a kick
velocity (such as in the case of binaries), then the merger sites
could have a substantial offset from the place of birth. The amount of
the offset will generally depend on a combination of the merger time
and the velocity of the binary system, as well as the potential well
in which the systems move.  Kick velocities are estimated to be in the
range of several tens to several hundreds km/s \cite{hobb05}. A binary
with a velocity in the upper range of the distribution, and a long
lifetime $\sim$ a few Gpc, can travel a substantial distance, on the
order of the Mpc scale, before merging. On the other hand, for
binaries with a short lifetime, on the order of a million years or
less, the place of merger is always going to be close to the place of
birth even for the fastest pulsars of the distribution. Therefore, it
is clear that the expectations for the distribution of the merger
sites are going to heavily depend on the expected distribution of
merger times of the binary population under consideration.

Another element of consideration is the fact that different types of
galaxies are expected to host different types of stellar populations,
with, at one extreme, starburst galaxies which are dominated by very
young stars, and at the other end, elliptical galaxies which have
almost no star formation rate at all and hence are dominated by an old
stellar population.  These different types of galaxies, therefore,
will select out different components of the binary mergers. Starburst
galaxies are expected to be dominated by the short-lived binary
population, while elliptical galaxies will mostly host the merger
events of the long-lived population. Spiral galaxies are in between,
since they host both young and hold star populations.  As a result of
these differences, merger events taking place in elliptical galaxies
will generally occur at larger offsets (from the galaxy centers) with
respect to merger events that occur in starburst and spiral
galaxies. Furthermore, within a given morphological galaxy type,
mergers in small galaxies will have substantially larger offsets than
merger in large galaxies. This is due to the larger gravitational
potential of the larger galaxies, which prevents the binaries from
traveling too far.

The sites of the binary mergers are especially important in determining
the observability of the GRB afterglows. Since the gas density declines
with the distance from the galaxy center, mergers that occur in the
outskirts of galaxies are expected to give rise to very dim afterglows.
Based on the considerations above, short bursts occurring in large starburst
galaxies are expected to be the brightest, while short bursts occurring
in small ellipticals will be generally the dimmest. A fraction of these
bursts are expected to be ``naked'', i.e. completely lacking any afterglow
emission, especially at wavelengths longer than the X-ray band, where the
effect of the density becomes more important. 

As more data on short bursts gather, the information on the fraction
of bursts as a function of the galaxy type will become an important
element of study in order to establish whether there is a dominant
channel of binaries (i.e. short-lived versus long-lived) that give
rise to short bursts. However, if one wants to extract from these types
of statistical studies meaningful physical information, one needs
to keep in mind that there is a huge potential for selection biases.
In fact, since bursts are localized through afterglow observations, and 
bursts occurring in elliptical galaxies are expected to be generally dimmer, 
there is going to be a selection effect toward the observation of a relatively
larger fraction of the bursts occurring in starburst and spiral galaxies,
i.e. the short-lived component of the binary population. Furthermore,
since the mean redshift of merger events for the short-lived population is 
higher than that for the long-lived population, the selection effect described
above will also result in a bias toward higher redshifts.

\subsection{The final moments: from a merging binary to a hyperaccreting
disk around a black hole}

Although there is no ``direct'' evidence for an accretion disk in
GRBs, the GRB phenomenology provides strong hints in that
direction. Firstly, accretion disks are a powerful way to tap
gravitational energy and channel it into other sources. Second, the
overall (short) burst durations, as discussed above, are naturally
accounted by the viscous timescale of the disk, while the millisecond
timescale variability (observed both in long and short bursts) is on
the same order of the dynamical timescale of a compact disk accreting
around a stellar-mass black hole. From an observational point of view,
on the other hand, it is well known that other systems in nature
believed to be associated with black holes accreting from a disk
(i.e. active galactic nuclei, micro-quasars) are able to power
relativistic jets, for which we have strong evidence also in GRBs.

In the last several years, a number of groups have devoted a
substantial effort into simulations of NS-NS and NS-BH mergers, and
the resulting structure of the hyperaccreting accretion disk. The
early simulations \cite{davi94} were Newtonian, used a polytropic
equation of state, and did not include the effects of
neutrinos. These, on the other hand, were shown to represent a
substantial source of cooling in a number of semianalytical, 1D
calculations of hyperaccreting disks around black holes
\cite{nara01,dima02} .  Neutrino effects, together with more realistic
equations of state, were taken into account into later simulations
\cite{ross99,ross02,ruff98,ruff99}. These simulations, independently
of the numerical method used, found results in general agreement.
Although details differ depending on whether the initial progenitor is
a NS-NS or a NS-BH, some common features can be identified.  As the
binary members spiral in, within a few orbital periods the outcome is
the formation of a dense and thick, hot torus, of mass on the order of
$\sim 0.01 - 0.3M_\odot$ that accretes onto a stellar mass black hole.
In the case where the initial binary members are both neutron stars,
the black hole will be formed as a result of the accretion of mass
onto one of the NSs. Depending on the total initial mass of the binary
system, the collapse of the hypermassive NS into a BH can occur
promptly, or it can be delayed for an initial time during which the
star is supported by differential rotation. The accreting material, on
the other hand, is provided by the tidally disrupted debris of the NS.

The duration of the prompt GRB phase phase is set by the time during
which efficient accretion occurs. Given the observed $\gamma$-ray
luminosities ($\sim 10^{49}- 10^{50}$ erg/s assuming a beaming
correction of a factor $\sim 0.1$), and taking an efficiency of
conversion of accretion energy into $\gamma$-rays of a fraction of
percent to a percent, the accretion rates of the disk must then be in
the range $\sim 0.01-10 M_\odot$. The resulting hyperaccreting disk is
very dense and hot, optically thick to photons, and cools mainly by
neutrinos.  In the upper end of the range of accretion rates, the
density of the disk can however be so high that the innermost regions
would become optically thick even to the neutrinos themselves.

An important component of the studies of GRB accretion disks deals
with the processes by which the disk black-hole system is able to
collimate and launch the relativistic jets known to power the GRBs.
Two mechanisms have been suggested as being involved: neutrino-anti
neutrino annihilation, and magnetic fields. In the former process,
suggested by a number of authors
\cite{good87,eich89,nara92,mesz92,moch95}, the source of energy is
provided by the neutrinos and anti-neutrinos emitted in the cooling
disk which annihilate in a funnel above a disk (which has a lower
density). Calculations \cite{dima00} have estimated that the maximum
efficiency by which the rest mass energy of the accreting material is
converted into neutrino luminosity does not exceed a value of $\sim
10^{-4}$. For a disk mass of $\sim 0.1 M_\odot$, this would yield an
energy output of about $10^{49}$ erg, therefore making this
jet-production mechanisms viable only if there is a substantial degree
of collimation in short bursts.

The second method of jet collimation and launch relies on the help of
magnetic fields. A number of authors \cite{nasa92,thom94,lyut06} have
suggesting this possibility by noticing how, even if the magnetic field
is initially low, it is likely to be amplified by the
magneto-rotational instability within the disk
\cite{balb91}. Numerical simulations \cite{mcki04} find that
magnetically driven jets, whose energy output increases with the spin
of the BH, are generally more efficient than neutrino-powered
jets. Magnetic fields are therefore considered to play an important
role in collimating and driving the jets from the accretion disk.

Whereas a substantial progress has been made in this area of research
of GRB hyperaccreting accretion disks, recent, new observations with
{\em Swift} have shown that the current picture is far from being
complete. In particular, the detection of energetic X-ray flares
superimposed on the smooth afterglow decay, with arrival times and
durations from tens to tens of thousands of seconds, requires the
presence of an engine with duration much longer than the fraction of a
second that is sufficient to power the prompt phase of a short burst
\cite{lazz07}. These observations have prompted a number of
suggestions on how to make the lifetime of the accretion disk much
longer than its viscous timescale. Perna, Armitage \& Zhang
(\cite{pern06}; see also Piro \& Phfal \cite{piro06}) suggested that
fragmentation of the outer parts of the accretion disk (which would
then accrete at later times) could be responsible for creating a
long-lived engine. Proga \& Zhang \cite{prog06}, on the other hand,
envisaged a scenario in which the accumulation of magnetic flux in the
innermost parts of the accretion disk creates a barrier that can then
produce intermittent accretion.  Other suggestions, which do not
involve accretion disks, include that of Dai et al. \cite{dai06}. They
showed that, if the NS-NS merger leads to a differentially rotating,
millisecond pulsar, then the differential rotation can cause the
interior magnetic field to wind up and break through the stellar
surface, hence resulting in magnetic reconnection-driven explosive
events. These events would be observed as X-ray flares.

\section{Gravitational waves from short GRBs}

Mergers of two compact objects have traditionally been of great
interest as sources of gravitational waves. With the likely
association of binary mergers with short GRBs, the interest of the
gravitational wave community has been extended to that of short
GRBs. The local rate of these sources, however, is not large enough to
make a detection likely with a blind search with LIGO-I. However, the
detection probability can be increased if the observations are made
shortly after the $\gamma$-ray detection from the burst is
detected. Estimates \cite{koch93} suggest that in this case a positive
detection in coincidence with a short GRB could be already made with
current gravitational wave detectors.  Clearly, such a signal, besides
giving information on the last moments of the binary merger, would
also provide the still needed conclusive evidence of the association
of short bursts with mergers of compact objects.

\index{paragraph}

%%%%%%%%%%%%%%%%%%%%%%%%%%%%%%%%%%%%%%%%%%%%%%%%%%%%%%%%%%%%%%%%%%%%%%  }

%%%%%%%%%%%%%%%%%%%%%%%%%%%%%%%%%%%%%%%%%%%%%%%%%%%%%%%%%%%%%%%%%%%%%%

\printindex
\end{document}